\documentclass[submission, Phys]{SciPost}
\usepackage{bm}

\begin{document}

\begin{center}{\Large \textbf {Photon pumping, photodissociation and dissipation at interplay for
the fluorescence of a molecule in a cavity}}\end{center}

\begin{center}
M. Gopalakrishna\textsuperscript{1},
E. Vi\~nas Bostr\"om\textsuperscript{2},
C. Verdozzi\textsuperscript{3*}
\end{center}

\begin{center}
{\bf 1} Department of Physics, Division of Mathematical Physics, Lund University, 22100  Lund, Sweden
\\
{\bf 2} Max Planck Institute for the Structure and Dynamics of Matter, Luruper Chaussee 149, 22761 Hamburg, Germany
\\
{\bf 3} Department of Physics, Division of Mathematical Physics and ETSF, Lund University, 22100  Lund, Sweden
\\
* Claudio.Verdozzi@teorfys.lu.se
\end{center}

\begin{center}
\today
\end{center}


\section*{Abstract}
{\bf
We introduce a model description of a diatomic molecule in an optical cavity, with pump and fluorescent fields, and electron and nuclear motion are treated on equal footing and exactly. The model accounts for several optical response temporal scenarios: a Mollow spectrum hindered by electron correlations, a competition of harmonic generation and molecular dissociation, a dependence of fluorescence on photon pumping rate and dissipation. It is thus a general and flexible template for insight into experiments where quantum photon confinement, leakage, nuclear motion and electronic correlations are at interplay.}

\vspace{10pt}
\noindent\rule{\textwidth}{1pt}
\tableofcontents\thispagestyle{fancy}
\noindent\rule{\textwidth}{1pt}
\vspace{10pt}

\section{Introduction}
\label{sec:intro}
Second harmonic generation (SHG) is the conversion by some material system of two photons of frequency $\omega$ into a single photon of frequency $2\omega$.  A classic hallmark of nonlinear optical behavior~\cite{bloembergen}, SHG still is, sixty years after its discovery~\cite{1stExpSHG}, the focus of extensive research in physics~\cite{physi}, engineering~\cite{engine}, chemistry~\cite{chemist}, biology~\cite{biolog}, and medicine~\cite{medicine}. Part of this interest stems from technology \cite{nloptdev,tech2}: SHG is the operating mechanism in optical devices and imaging techniques that are surface or interface sensitive~\cite{Shen,Shen1,McGilp}. Another reason is that there are aspects and regimes of SHG still not fully understood, making it a valuable benchmark for advances in nonlinear optics.

Several theoretical methods are used to describe SHG~\cite{Myrta}, from nonlinear response in frequency space~\cite{Luppi} to  
Bloch-Maxwell equations~\cite{Hughes} and real-time first-principle approaches~\cite{TDDFTSHG,Claudio1,Myrta,ReviewHHG,RubioHHG1}. Often, classical radiation fields are used, which is appropriate in the strong field limit. However, highly interesting effects in SHG (and fluorescence in general) appear in the low photon regime~\cite{QuantumSHG,QSHG1,Cini93,Cini95}, where quantum effects generally dominate \cite{Camacho2021} and the so-called rotating wave approximation (RWA)~\cite{JaynCum,CarmaRWA,JaynCumReview,K.Fujii,QXie2017} may be inadequate~\cite{PerfettoMollow,OBrien,IOP2017}.

Optical cavities permit an accurate selection of confined electromagnetic modes~\cite{Schleich,cavity,CavityReview}, and
allow to address the low photon regime of SHG~\cite{BACV}. However, key elements left out of many theoretical works on few-level systems is an explicit description of electronic correlations and nuclear dynamics, even though these
can importantly affect the harmonic signal~\cite{SHG_corr,SHGphonon1,PhotoDiss}. First-principle
descriptions include these contributions~\cite{Myrta,Luppi,TDDFTSHG,RubioHHG1}, but usually approximations are made in numerical implementations. Therefore, because of the broad relevance of SHG, it is useful to consider model systems where photon pumping, cavity leakage, electronic correlations, and nuclear motion can be treated exactly and on equal footing, to gain a generic and accurate understanding of their interplay.

In this work we introduce a simple and flexible theoretical framework to describe a single molecule embedded in an optical cavity, and study its fluorescence properties. Within this framework all the aforementioned effects and interactions are considered, and the following picture emerges: $(1)$ the SHG signal is larger for faster photon pumping; $(2)$ electron-electron interactions strongly reduce the fluorescence signal; $(3)$ for light atomic masses photodissociation takes place, inhibiting  fluorescence and SHG; for heavier masses, the opposite occurs; $(4)$ both resonant and SHG signals are quenched in time by cavity leakage. While not tied to any specific molecule, our results unveil a multifaceted light-matter scenario for SHG and fluorescence in the low photon regime, when multi-photon effects are important. At the same time, they give qualitative but rigorous initial insight for more refined investigations of systems of direct experimental interest.

\section{Hamiltonian, initial state and fluorescent spectrum}
We consider a homo-nuclear diatomic molecule embedded in a cavity, where each atom has a mass $M$ and a single $s$-orbital. The molecule is occupied by two electrons of opposite spin, interacting with a cavity field of frequency $\omega_0$ and an fluorescent field of frequency $\omega$. The molecule and cavity are assumed to be one-dimensional, with the molecular axis aligned with the axis of the cavity. The total Hamiltonian reads $\hat{H}(t)=\hat{H}_s(t)+ \hat{V}_{\rm ext}(t)$, where the system Hamiltonian is $\hat{H}_s(t)=\hat{H}_{\rm mol}+\hat{H}_{\rm rad}+ \hat{H}_{\rm int}(t)$ and $\hat{H}_{\rm mol}$, $\hat{H}_{\rm rad}$ and $\hat{H}_{\rm int}(t)$ respectively describe the molecule, the photon fields, and the light-matter interaction ~\cite{Schleich}. The external field term, $\hat{V}_{\rm ext}(t)$, will be specified later. In more detail, the molecular Hamiltonian we use is
\begin{align}
 H_{\rm mol} = \frac{\hat{P}^2}{2(2M)} + \frac{\hat{p}^2}{2(M/2)} + \frac{C}{\hat{x}^4} + U\sum_i\hat{n}_{i\uparrow}\hat{n}_{i\downarrow} \nonumber 
  - Ve^{-\lambda \hat{x}}\sum_{\sigma}(c_{1\sigma}^\dagger c_{2\sigma}+c_{2\sigma}^\dagger c_{1\sigma}), \label{H_m}
\end{align}
where the first two terms give the kinetic energy of the molecular center of mass (with momentum $\hat{P}$), and relative atomic motion (with momentum $\vec{p}$). The third term accounts for an inter-atomic repulsion of strength $C$, $\hat{x}$ the inter-atomic coordinate. Finally, the remaining terms of $\hat{H}_{\rm mol}$ describe the electron dynamics via an intra-orbital repulsive interaction of strength $U$, and a kinetic energy term arising from electrons hopping between the atoms. Here $\hat{n}_{i\sigma} = c^\dagger_{i\sigma}c_{i\sigma}$ and $c^\dagger_{i\sigma}$ creates an electron with spin projection $\sigma$ at atom $i$. The strength of this term is proportional to $V$, but it also depends on the internuclear distance via the operator $e^{-\lambda \hat{x}}$ (with $\lambda$ an attenuation parameter). This gives a phenomenological 
(but intuitively physically plausible  \cite{pap1,pap2,pap3,pap4}) fully quantum mechanical interaction between the electrons and the inter-atomic motion. In the numerical calculations, we set  $V = -2$, $C = 0.6$ and  $\lambda = 0.6$, to obtain a Morse-like potential landscape for inter-atomic motion, and an equilibrium position $r_0 = 1.156$. In this way, the effective hopping $V_{\rm eff} = V\exp(-\lambda r_0) \approx -1$ in equilibrium.
  
The second contribution to $\hat{H}_s$ describes the two photon modes, $\hat{H}_{\rm rad} = \omega_0 b^\dagger b + \omega b'^{\dagger}b'$, with $b$ ($b'$) destroying a cavity (fluorescent) photon with frequency $\omega_0$ ($\omega$). For computational simplicity we exclude the direct interaction between modes and nuclei, and neglect center of mass motion~\cite{semiclassical}. The cavity-molecule interaction is thus $\hat{H}_{\rm int} = \hat{M} \big[g_c (b^\dagger+b)+g'(t)(b'^\dagger+b') \big]$, where $\hat{M} = \sum_\sigma (c_{b\sigma}^\dagger c_{a\sigma} + c_{a\sigma}^\dagger c_{b\sigma})$ and $c_{b/a} = (c_{1} \pm c_{2})/\sqrt{2}$ destroys an electron in the molecule's bonding or antibonding state.  In the calculations, the fluorescent coupling is damped, i.e. $g'(t) = g_f\exp(-\Gamma t)$ (we set $\Gamma = 0.02$), to describe phenomenologically cavity losses~\cite{Cini93,BACV}. We will also consider a more rigorous description of cavity leakage, by coupling the system to a bath of harmonic oscillators. Since the temporal change of $\hat{V}_{\rm ext}(t)$ is restricted to a short initial fraction of the simulation interval, $\hat{H}(t)$ and $\hat{H}_s(t)$ are  time-independent at long times.
We will consider two initial light+matter states: i) A product state $|\Psi'_0 \rangle\equiv |g_m\rangle |\beta\rangle_c |0\rangle_f$, with the molecule in its ground state $|g_m\rangle$ for $g_c=g_f=0$, the cavity field in a coherent state $|\beta\rangle_c$, and the fluorescence field in its vacuum state $|0\rangle_f$. ii) The ground state $|\Psi''_0\rangle\equiv |g\rangle$ of the full Hamiltonian $\hat{H}_s(t=0)$.

\subsection{Resonance frequency and fluorescence spectrum}
\label{sec:another}
We consider a cavity mode with a frequency of either $\omega_0 = \Omega_R$ in resonance with the molecule's electronic transitions, or $\omega_0 = \Omega_R/2$. Due space and spin symmetries, the molecule's electronic ground state is a spin singlet of even parity. Since the 
total electron spin $S$ is conserved in absorption and emission,  $\Omega_R = E^{\rm ex}_{{\rm odd},S=0}-E^{g}_{{\rm even},S=0} = U/2 + [4V_{\rm eff}^2 + (U/2)^2]^{1/2}$~\cite{otherpresc} (see
Appendix \ref{APPA}). Concerning the value chosen for the interaction among the electrons, in Appendix \ref{APPB} we show that fluorescence weakens on increasing the electronic correlations. Accordingly, in the rest of the paper we focus on the weakly interacting regime where $U = 1.0$ and $\Omega_R = 2.56$.

We characterize the fluorescence spectrum in terms of
\begin{equation}
\mathcal{P}(\omega',t)=\sum_{\lambda r_i n} \sum_{m>0}\rvert \langle \lambda r_i nm|\mathcal{T}\big[e^{-i\int_0^t\hat{H}(t')dt'}]\big|\Psi_0\rangle \rvert^2,
\end{equation}
where $\mathcal{P}$ is the probability  to have one or more photons in the fluorescence mode $\omega'$ at time $t$ \cite{Cini93}.
Here $|\Psi_0\rangle$ is a given initial state (i.e., either $|\Psi'_0\rangle$ or $|\Psi''_0\rangle$ above)  and the $\omega'$-dependence is contained in $\hat{H}(t)$. The sums over $\lambda$, $r_i$ and $n$ trace out electronic, nuclear and cavity mode degrees of freedom, while the sum over $m$ ensures that at least one fluorescent photon is emitted. The real-time dynamics of the system (with coupled electronic, atomic and photonic degrees of freedom) was obtained via the short iterated Lanczos algorithm, by computing the exact time evolved many-body state $|\Psi(t) \rangle$ starting from $|\Psi_0\rangle$. The configuration size of the problem is $N = 4N_c N_f N_R$, where $4$ is the dimension of the electronic subspace, and $N_c$, $N_f$, and $N_R$ are respectively the maximum number of cavity photons, fluorescence photons, and grid points for the nuclear coordinate $x$. We have ensured numerical convergence with respect to these parameters.

\begin{figure}
\includegraphics[width=1.0\columnwidth]{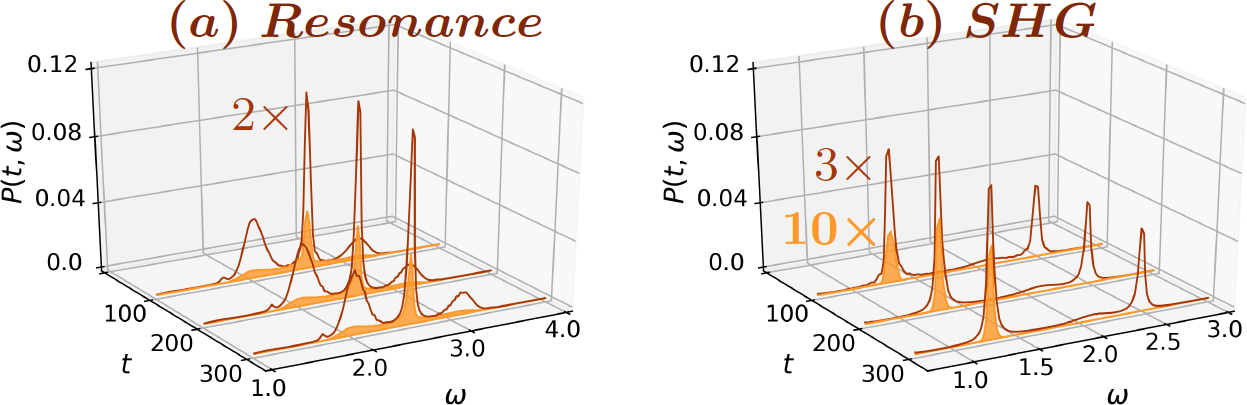}
 \caption{(a) Resonant response for $\omega_0 = \Omega_R$ and (b) SHG response for $\omega_0 = \Omega_R/2$ of a rigid molecule, starting from a coherent state $|\Psi'_0\rangle$ with $\beta^2 = 9$ (empty curves) and from the cavity+molecule's ground state $|\Psi''_0\rangle$ followed by pumping (filled curves). For the pumped cavity, the drive is kept on until $\langle b^\dagger b \rangle \approx 9$ $t_1=\frac{6\pi}{\omega_0}$ and $t_2=\frac{31\pi}{\omega_0}$, with $g_d = 0.229$ and 0.0996 in (a) and (b) respectively.  In all panels $U = 1.0$, $g_c = 0.08$, $g_f = 0.01$ and $\Omega_R = 2.56$. Plots are scaled for visual clarity and the scaling factors are indicated in color.}
\label{fig1}
\end{figure}

%
\section{Fluorescence in a rigid molecule and initial state preparation} 
In a cavity with low photon number,  SHG is remarkably sensitive to the system's initial state. This important point is illustrated by comparing the spectra resulting from the different initial states $|\Psi'_0\rangle$ and $|\Psi''_0\rangle$ introduced earlier. With $|\Psi'_0\rangle$, which is a coherent state with $\beta^2$ photons and not an eigenstate of $\hat{H}_s(t)$, the system evolves under the full Hamiltonian $\hat{H}_s(t)$ and $\hat{V}_{\rm ext} = 0$. Thus, fluorescence photons are emitted in time. For $|\Psi''_0\rangle$, and with the parameters we consider, the initial occupation of the cavity mode is negligible ($ < 10^{-3}$). So, for a meaningful comparison with the results from $|\Psi'_0\rangle$, the cavity is pumped by a driving field of frequency $\omega_0$, until an approximately coherent state with average photon number $\langle b^\dagger b\rangle \approx \beta^2$ is reached \cite{window}. 
\begin{figure}[t]
\begin{center}
 \includegraphics[width=0.8\columnwidth]{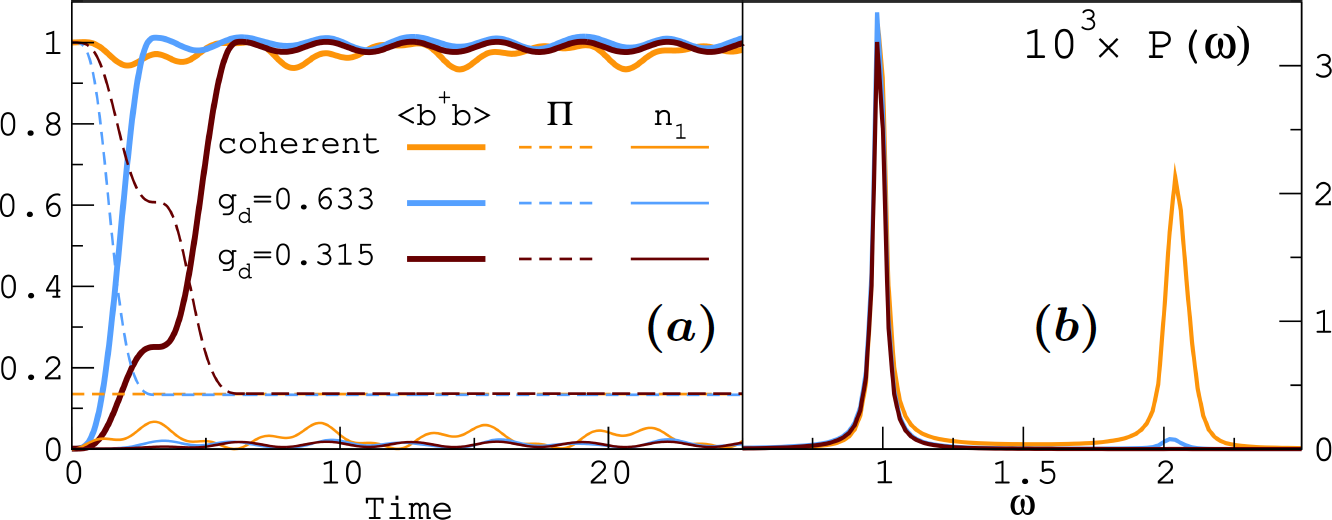}
 \end{center}
 \caption{Cavity pumping in a two level system 
 with $g_c = 0.1, g_f = 0.01, \Gamma = 0.02$, and $\omega_0 = \Omega_R/2=1$. 
Starting from the same ground state, two pumping speeds are considered with $t_s=\frac{\pi}{\omega_0}$ and $t_s=\frac{2\pi}{\omega_0}$ respectively. Reference results 
from an initial coherent state ($\beta^2= 1$) and no pumping are also shown. (a) Time-evolved average number of cavity photons, total parity and excited state population. (b) Corresponding SHG spectra at long times.}
 \label{fig2}
\end{figure}
The spectra for the two initial configurations, and the low photon limit $\beta = 3$~\cite{footnote}
are in Fig.~\ref{fig1}, for both the resonant ($\omega_0 = \Omega_R$) and SHG ($\omega_0 = \Omega_R/2$) cases. In the resonant case, and starting from $|\Psi'_0\rangle$ (Fig.~\ref{fig1}a), a spectrum with well-defined Mollow features emerges already at early times and converges to a similar profile at longer times. These features can be understood from a dressed-level picture~\cite{Cini93,BACV} since the cavity mode is in resonance with a parity allowed transition. Interestingly, starting from $|\Psi''_0\rangle$ and pumping the cavity up to $\beta = 3$ (Fig.~\ref{fig1}b), the spectrum at long times is qualitatively similar to Fig.~\ref{fig1}a, although the intensity of the Mollow sidebands is reduced compared to the main peak. 
A markedly different picture emerges in the SHG regime: For initial state $|\Psi'_0\rangle$ (Fig.~\ref{fig1}c), the spectrum quickly develops two sharp features (with a broad shoulder in the middle) corresponding to a Rayleigh (SHG) contribution at $\omega_0$ ($2\omega_0$). However, when starting from the full ground state $|\Psi''_0\rangle$ and pumping the cavity, the SHG signal is strongly suppressed at all times (Fig.~\ref{fig1}d). 

\subsection{The dependence on the initial conditions} The rationale for the above results is that 
the SHG signal strongly depends on the pumping rate. To uphold our statement, we consider
for simplicity
SHG in a two-level system (TLS) with levels $|0\rangle$ and $|1\rangle$ and $\omega_0 = \Omega_R/2$. In Fig.~\ref{fig2}a we show the evolution of the total parity $\Pi = \langle e^{i\pi b^\dagger b} (\hat{n}_0-\hat{n}_1)e^{i\pi b'^\dagger b'}\rangle $, the cavity mode occupation, and the occupation $n_1$ of the TLS excited state. The dynamics is obtained starting either from a product state with the cavity mode in a coherent state (with $\beta^2 = 1$), or from the exact ground state where the cavity mode is pumped at different speeds until $\langle b^\dagger b\rangle \approx 1$. 

Fig.~\ref{fig2}b shows the corresponding long-time limit SHG. When starting from $|\Psi'_0\rangle$, $\Pi$ has a constant mixed parity $\Pi_{\rm coh} \approx 0.17$. By contrast, when starting from $|\Psi''_0\rangle$, initially $\Pi$ is 1, but then drops to $\Pi_{\rm coh}$ with pumping.  Thus, in both cases and at almost all times, the system has mixed parity (which is necessary for SHG in a TLS~\cite{BACV}). Yet, the SHG signal is absent for slow pumping and very small for fast ramping. Further insight comes from how the population $n_1$ of the excited level changes in time: it is very small for the pumped cases, but noticeably large for the coherent case. Thus, the cavity pumping speed strongly affects the population of the excited level and the SHG strength, which increases for faster drives, and similar trends are observed for the resonant regime (see Appendix \ref{APPC}). While exemplified for a TLS, our considerations equally hold  for the molecule investigated in the rest of the paper.

\begin{figure}[t]
\includegraphics[width=1.0\columnwidth]{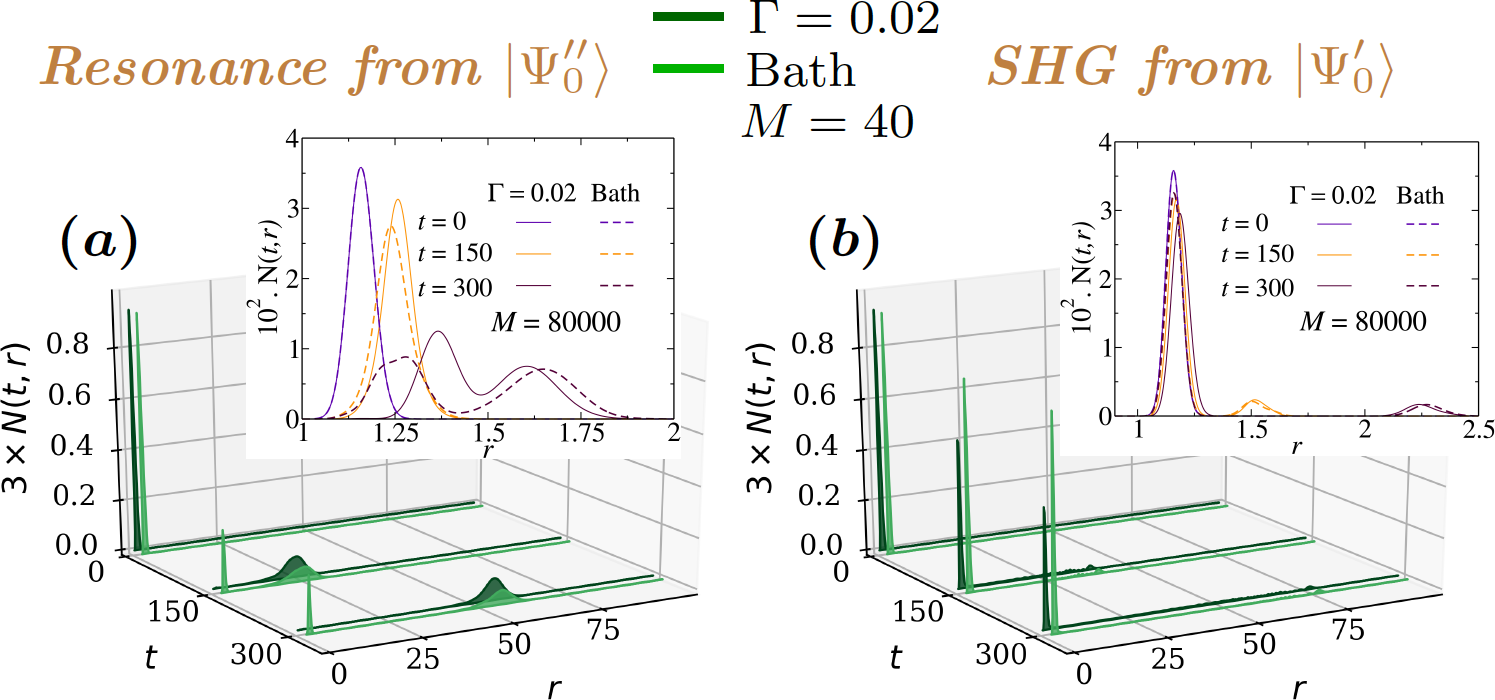}
\caption{Dynamics of the relative interatomic distance in resonant (a) and SHG (b) regimes, for atomic masses $M = 40$ (main plots) and $M = 8\times 10^4$ (insets). In all panels $U = 1$ and $r_0 = 1.156$. At resonance (a) the calculations were performed by pumping the cavity until $\langle b^\dagger b \rangle \approx 9$ starting from the interacting ground state $|\Psi''_0\rangle$, with $g_c = 0.03$, $g_f = 0.01$ and $g_d = 0.151$. For SHG (b) the calculations started from the product state $|\Psi'_0\rangle$ with the cavity field in a coherent state with $\beta^2 = 9$, with $g_c = 0.08$, $g_f = 0.01$ and $\omega_0 = 1.28$. In all panels cavity dissipation is described either via exponential damping ($\Gamma = 0.02$ curves), or via coupling to a bath of classical oscillators (``Bath'' curves) with $C_k = A (\Delta k)^a$, $N_B = 1000$ oscillators, $A = 0.01$, $a = 0.6$ and $\Delta k = 0.01$.} 
 \label{fig3}
\end{figure}

\section{Cavity leakage and atomic motion} For a more microscopic treatment of the cavity leakage, we now set $g'(t) = g_f$, and couple both photon modes to a bath of independent oscillators, described by the Hamiltonian $\hat{H}_{\rm bath}= (1/2) \sum_{k=1}^{N_B} (\hat{p}_k^2+ \omega_k^2 \hat{x}_k^2 )$. The coupling of bath and cavity modes is of the Caldeira-Leggett type~\cite{Caldeira,Venkataraman,Hermann}, i.e. 
$\hat{H}_{\rm diss}=-\sum_{k=1}^{N_B} C_k \hat{x}_k [(b^\dagger+b)+(b'^\dagger+b')]$, and the distribution of the oscillators is determined by the density of states $J(\omega)=\sum_{k=1}^{N_B} (C_{k}^2/\omega_k) \delta(\omega-\omega_k)$. In the actual calculations $\omega_k = k\Delta$ and $C_k = A \omega^a$. The values of $N_B$, $A$, $\Delta$ and $a$ determine the decay rate of the photons (the cavity quality). The bath variables are propagated via Ehrenfest dynamics, $\ddot{x}_k(t) = -\omega_k^2 x_k(t) + C_k [\langle b^\dagger + b\rangle_{\bar{x},t} + \langle b'^\dagger + b'\rangle_{\bar{x},t}]$, where $\bar{x}\equiv \{x_{k}\}$. In turn, the coordinates $\bar{x}$ enter parametrically into $|\Psi(t)\rangle$. While computationally inexpensive,
this treatment of the bath keeps the dynamics unitary and Hermitian (see Appendix \ref{APPD} for further details) 

\begin{figure}[t]
\begin{center}
\includegraphics[width=0.8\columnwidth]{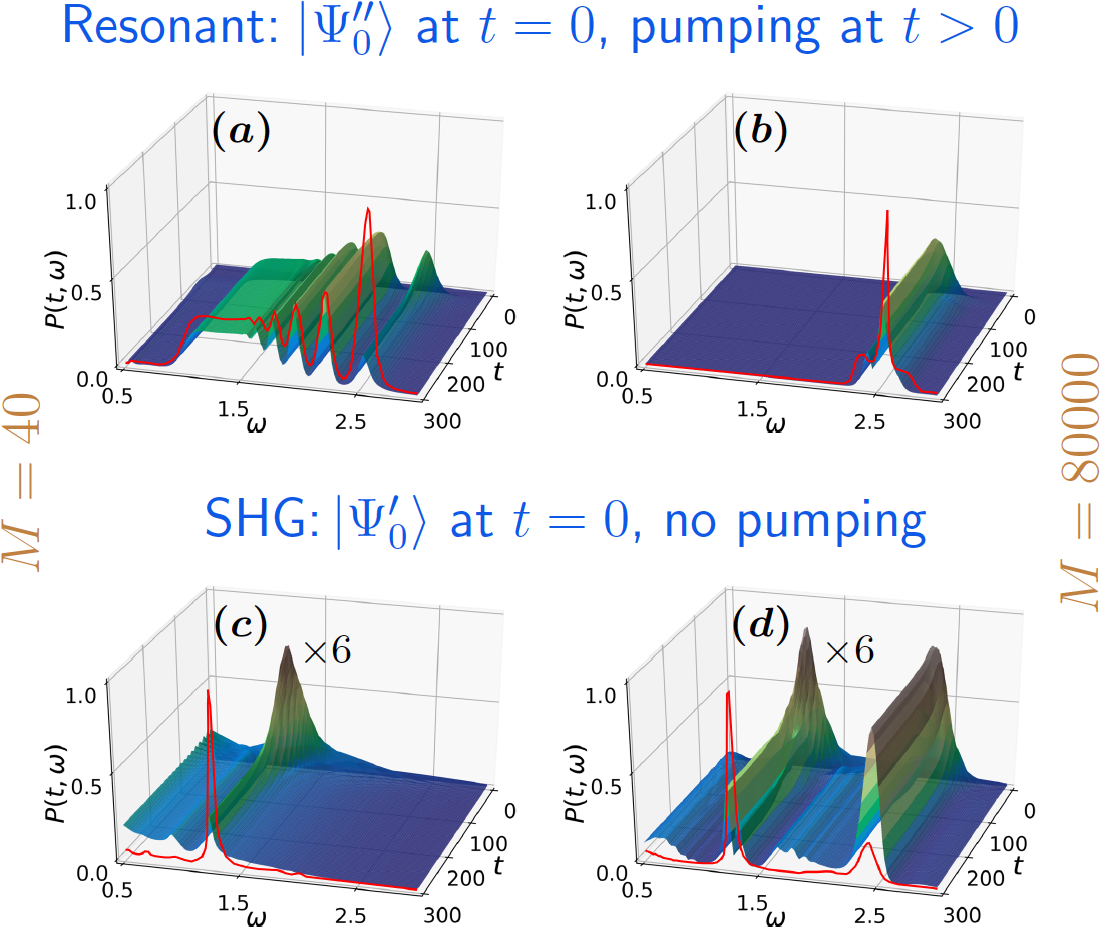}
\end{center}
\caption{Time-dependent fluorescence for atomic masses $M = 40$ (a ,c) and $M = 8\times 10^4$ (b, d). The time evolution was performed with a bath of $N_B$= $1000$ oscillators, with $C_k = A (\Delta k)^a$, $A=0.01$, $a=0.6$ and $\Delta k=0.01$. The red curves show long-time limit of $\mathcal{P}(\omega',t)$ for an exponential dissipation with $\Gamma = 0.02$. (a,b) Resonant case, starting from $|\Psi''_0\rangle$ and pumping the cavity until and 
 $\langle b^\dagger b\rangle \approx 9$, with $t_1=\frac{6\pi}{\omega_0}$, $t_2=\frac{41\pi}{\omega_0}$, $g_c = 0.03$, $g_f = 0.01$,  $\omega_0 = 2.56$, $g_d = 0.151$. (c,d) SHG case, starting from $|\Psi'_0\rangle$ with $\beta^2 = 9$, $g_c = 0.08$, $g_f = 0.01$ and $\omega_0=1.28$. The time-evolved plots are magnified for visual clarity, and in all cases $U = 1$ and $r_0 = 1.156$.}
 \label{fig4}
\end{figure} 

\subsection{Nuclear motion} 
Until now, the molecule was kept rigid at interatomic distance $r_0$ corresponding
to the maximum of $N(r,t=0)$, the equilibrium probability distribution of the nuclear relative coordinate $r$.
How the interatomic distance is affected by the light-matter interaction (and viceversa) is shown in Fig.~\ref{fig3}, where we display  time snapshots of $N(r,t)$ for both resonant and SHG regimes.
We include cavity leakage via either exponential attenuation ($g'(t) = g_f e^{-\Gamma t}$) or the interaction with an oscillator bath.  In the resonant regime, the system is initially in its ground state $|\Psi''_0\rangle$ and the cavity mode is subsequently pumped. In this case, the molecule dissociates quite rapidly for $M=40$, irrespective of the type of damping process considered. Conversely, for the larger mass, no dissociation occurs in the simulation interval, and the atoms remain around the 
equilibrium configuration with a broadened distribution $N(r,t)$.

In the SHG regime, the system's initial state is $|\Psi'_0\rangle$ for both values of $M$. Here, the molecule predominantly remains close to the equilibrium configuration at all times, especially when field damping is described as an oscillator bath. The tendency to delocalise is enhanced in by an exponential damping, indicating that cavity leakage also plays a role. As shown next, the different atomic dynamics affect the optical response in distinct ways.

\section{Molecular dissociation and optical response} Fig.~\ref{fig4} shows the fluorescence spectra for finite $M$, with all the elements previously discussed (photon pumping speed, atomic dynamics and cavity leakage) at interplay. The spectra in panels (a,b) and (c,d) respectively correspond to the atomic probabilities $N(r,t)$ of Fig.~\ref{fig3}a and Fig.~\ref{fig3}b.
At resonance, the fluorescence spectrum strongly depends on the value of the atomic mass: For $M = 40$  the molecule dissociates (see Fig.~\ref{fig3}a) and $\mathcal{P}(\omega',t)$ exhibits sharp features as well as a plateau, in stark difference to the Mollow-like structure of the rigid molecule limit. Conversely, for $M = 8\times 10^4$, the molecule remains localized around the equilibrium position (inset in Fig.~\ref{fig3}a), and at long times $\mathcal{P}(\omega',t)$ is peaked around the resonant value $(\Omega_R = 2.56$). Overall, the shape of $\mathcal{P}(\omega',t)$ for exponential and bath dissipation show a mutual resemblance at long times. However, for bath dissipation the intensity of $\mathcal{P}(\omega',t)$ is considerably weaker. This is clearly manifest in the large $M$ case, where a Mollow triplet is well defined for exponential damping but only partially reproduced (with less intensity) when the system evolves in the presence of an oscillator bath.

A quite different picture emerges for SHG regime (Fig.~\ref{fig4}c and d), where $\mathcal{P}(\omega',t)$ is considerably weaker in the case of an oscillator bath. Also, when the molecule dissociates (Fig.~\ref{fig4}c), the SHG signal is absent irrespective of the type of dissipation. Conversely, for larger $M$, the SHG signal is present if the system evolves in contact with an oscillator bath, but with smaller intensity than for exponential dissipation. This suggests that the multi-photon cavity field is much more affected by dissipation under off-resonant conditions than at resonance. This picture persists also when considering 
the effect of the driving field strength for different form of dissipation (see Appendix \ref{APPE}).

In summary, in the dissociation regime both resonant Mollow and SHG signals are quenched. Also, for dissipation via an oscillator bath, for a broad range of atomic mass values fluorescence is always vastly reduced. Finally, even with no cavity leakage, the strength of the SHG response is determined by the cavity pumping rate.

\section{Conclusion}
Many decades of nonlinear optics research gave us a robust conceptual understanding of SHG, and actual uses in technology. Yet, some SHG regimes remain little explored, and how different physical mechanisms and interactions contribute to fluorescence is not always understood. In this work, we studied theoretically one of these (namely, the low photon) regimes, using a model molecule in an optical cavity, and via an exact time-dependent configuration interaction (TDCI) approach, where all quantum degrees of freedom (electrons, photons and relative atomic motion) are included on equal footing and supplemented by a semi-classical treatment of cavity dissipation. 

Our study reveals a previously unknown, complex landscape for fluorescence, where the latter
is reduced by electronic interactions and by cavity leakage, enhanced by fast cavity pumping, and quenched by molecular photodissociation. These competing trends likely occur in real molecules as well; it should thus be possible to detect them in experiments at low photon regimes. 
Our theoretical and computational framework can be applied and extended in different ways,
e.g.more realistic molecules, or cavities with more than one molecule. Other possibilities
are few ultracold bosons in cavities, to provide
insight for SHG in the Gross-Pitaevskii limit, or fermions in the (interacting) Dicke's model, 
in conjunction with other techniques that  exhibit better size-scaling behavior than TDCI, 
e.g. nonequilibrium Green's functions. 
Some of these undertakings are under way.

\section*{Acknowledgements}
We acknowledge A. D'Andrea for discussions. 
\paragraph{Author contributions}
M.G. performed all calculations and interpretation of results under the supervision of E.V.B. and C.V. 
The project was conceived  by E.V.B. and C.V. The overall supervision of the project was by C.V. Both 
M.G. and E.V.B. contributed to the writing of the code. All authors collaborated in writing the paper.
\paragraph{Funding information}
M.G. and C.V. acknowledge support from the Swedish Research Council (grant number 2017-03945).

\begin{appendix}

\section{Further details and additional results}\label{APPA0}
\subsection{Resonant frequency for the dimer molecule}\label{APPA}
To discuss the selection rules for light absorption, it suffices to consider a fixed molecule. The Hamiltonian is
\begin{align}
H_e=-t \sum_{\sigma}(\hat{c}^\dagger_{1\sigma}\hat{c}_{2\sigma}+ \hat{c}^\dagger_{2\sigma}\hat{c}_{1\sigma}) + U \sum_{i=1,2} \hat{n}_{i+} \hat{n}_{i-} \label{dimer},
\end{align}
where $t>0$. The molecule-light interaction for the two cavity modes is taken  as 
\begin{align}
H_{int}&= \bigg[\sum_{\sigma}g_{1}(\hat{b}^\dagger_\sigma \hat{a}_\sigma+\hat{a}^\dagger_\sigma\hat{b}_\sigma)\bigg]
(\beta_1^\dagger+\beta_1) + (1 \rightarrow 2)\nonumber\\
&\equiv \hat{M}_1 (\beta_1^\dagger+\beta_1)+\hat{M}_2 (\beta_2^\dagger+\beta_2).
\end{align}
For two electrons of opposite spin, $H_e$ has three singlet eigenstates ($S=S_z=0$) and one triplet eigenstate
($S=1,S_z=0$). The eigenvalues are 0 for $S=1$ and $U, U/2\mp \sqrt{4t^2+(U/2)^2}$ for $S=0$. The ground
state	is the singlet with energy $U/2 - \sqrt{4t^2+(U/2)^2}$, and it is even under spatial parity. The eigenstates with
odd symmetry under parity have energies $0$ with $S=0$ and $U$ with $S=1$.

It can be easily shown that 
optical transitions between the two even ($E$) many-body states or between the two odd ($O$) many-body  states are forbidden (e.g. $\langle E_1|\hat{M}_{1,2}|E_2 \rangle=0$), and the
only permitted transitions are between odd and even ones (i.e. with opposite parity). 
Furthermore, using the matrix expressions above for $\hat{M}_{1/2} $ and $\hat{\bf{S}}^2$, one can show that 
$[\hat{M}_{1/2},\hat{\bf{S}}^2] = 0$. So
the only transition allowed from the ground state
is the even-odd one where the system goes $|g,S=0\rangle \rightarrow |O, S=0\rangle$ and where the energy difference is
$\Omega_R= E_{O,S=0}-E_{g,S=0}= U/2+\sqrt{4t^2+(U/2)^2}$, 
which defines the``many-body'' resonance condition for the $\omega_1$ field
in perturbation theory, similar to the two-level single-particle case. More in general, for 
the multi-photon case of interest here, the bare electronic many-body levels are renormalised by the photons, parity gets mixed up, and more transitions are possible and, most importantly, the parity of the full electron+photon systems must be considered.
In the presence of nuclear dynamics, the values of the effective hopping parameter in the dimer changes in time and so it does  
$\Omega_R$.

 \subsection{The interaction parameters}\label{APPB}
\begin{figure}
\begin{center}
\includegraphics[width=0.5\columnwidth]{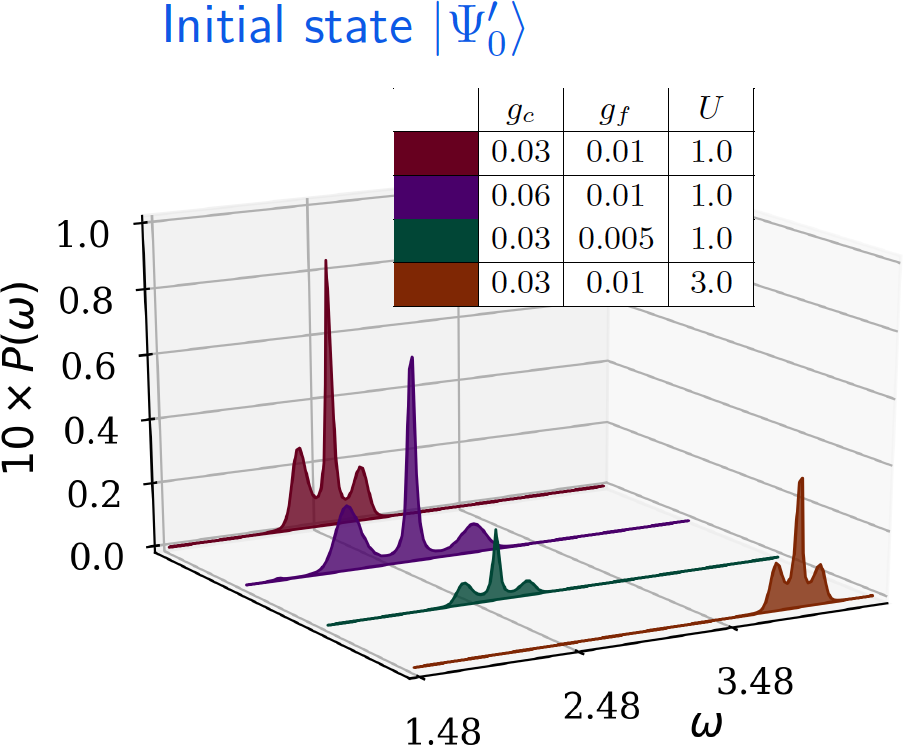}
\end{center}
	\caption{Fluorescent spectra for the rigid molecule starting with a coherent state with with  $\beta=3$, $r_0=1.156$, $\Gamma=0.02$, $\Omega_R=2.56$ and $\omega_0=\Omega_R$.    } 
	\label{figSM1}
\end{figure} 
Before choosing the values for the parameters $g_{c}$, $g_{f}$ and $U$ used in the paper, we have performed calculations to observe their effect on the spectra. A sample of the ensuing results is reported in Fig. \ref{figSM1}. Due to coupling between light and the molecule, the molecular levels will split and the splitting energy is $\propto g_{c}$ \cite{BACV}. Hence the regime of the emitted photon frequency will be affected by the incident field coupling, as observed in Fig. \ref{figSM1}. On increasing $g_{c}$, the fluorescent spectra get broadened, since this involves large range of frequencies for the emitted photon. On the other hand, Increasing the coupling $g_{f}$ increases the intensity of the fluorescent spectra. The electron interaction $U$ hinders electronic hopping between the two sites of the molecule. The emission of the fluorescent photon requires a transition among bonding and the anti-bonding molecular levels, and thus it involves electron hopping between the molecular sites. Accordingly, increasing the electron interaction decreases the intensity of the emitted photon, as it can be observed  in Fig.~\ref{figSM1}.    
%
\subsection{Pumping rate and resonant regime for a two-level system}\label{APPC}
In Fig.~\ref{fig2}, we show $\mathcal{P}(\omega)$ for $\omega_0 = \Omega_R$ for two driving speeds as well as for photons initially in a coherent state. We observe similar trends as in the SHG regime discussed in Fig. 2, namely fast pumping leads to closer agreement with the coherent state spectrum. Since photons interact with the TLS during the drive, the coherent and fast-drive spectra become increasingly similar when the system-cavity interaction $g_c$ is decreased.

\begin{figure}
\begin{center}
\includegraphics[width=1.0\columnwidth]{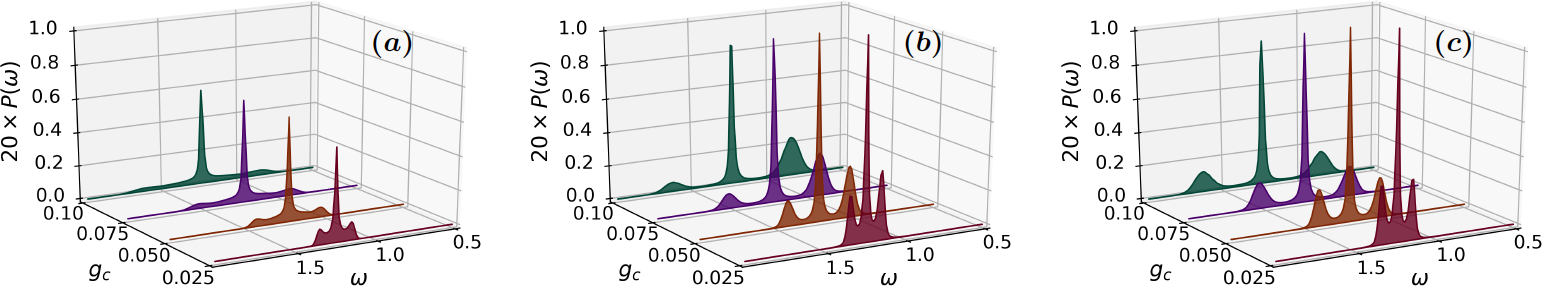}
\end{center}
\caption{
Long-time limit of fluorescence spectra for a two-level system in the resonant regime and in a pumped cavity with $g_d=0.232$, $t_1=\frac{6\pi}{\omega_0}$, $t_2=\frac{31\pi}{\omega_0}$, (a) and $g_d=5.150$, $t_s=\frac{\pi}{\omega_0}$ (b). The pumping is applied until $\left<b^\dagger b\right> \approx 16.0$ in  the cavity and the initial state is $|\Psi_0 ''\rangle$. Reference results
starting the time evolution from an initial coherent $|\Psi_0 '\rangle$ state with $\left<b^\dagger b\right> =16.0$ but without pumping are also
shown (c).  Spectral intensities are in arbitrary units, and parameters common to all panels are $g_f = 0.01$, $\Gamma = 0.02$ and $\omega_0 =\Omega_R= 2.0$.
} 
\label{fig2}
\end{figure}

\subsection{Cavity leakage via a Caldeira-Leggett bath: some details}\label{APPD}
To damp the driving field, 
we use ideas borrowed from the physics associated with the Caldeira-Leggett model CLM).
The CLM is defined as
\begin{equation}
H=\frac{p^2}{2M}+V(x)
+\sum_{k=1}^N \bigg[ \frac{p_k^2}{2m_k} +\frac{1}{2}m_k\omega_k^2 \bigg(x_k -\frac{C_k}{m_k \omega_k^2}x\bigg)^2\bigg].
\label{EqCLM}
\end{equation}
The classical treatment of Eq.~(\ref{EqCLM}) gives the solution 
\begin{equation}
M\ddot{x}(t)+M\int_{t_0}^t\gamma(t-t')\dot{x}(t')dt'=-M\gamma(t-t_0)x(t_0)+F_L(t),
\end{equation}
where $\gamma(t)$ determines the dissipative features of the bath
(for example, for $\gamma(t)\rightarrow \gamma_0\delta(t)$, we have a standard friction term).
Making use of Fourier/Laplace transforms, in the continuum-bath limit we get
\begin{equation}
\gamma(t)=\frac{2}{\pi}\int \frac{J(\omega)}{M\omega}\cos\omega t~d\omega, 
\end{equation} 
where 
$J(\omega)=\frac{\pi}{2}\sum_{k=1}^N \frac{C_k^2}{m_k\omega_k} \delta(\omega-\omega_k)$ is
the spectral density of the bath. Often, in practice, one takes $J(\omega) \propto \omega^\alpha$ 
in an interval range $[0,\omega_c]$, and zero otherwise. 
To describe dissipation/leaking for the cavity modes, we adopt a modified form of the CLM, 
where i) the cavity modes are in the second quantisation  
picture and ii) the requirement of translational invariance is neglected. Using here as example
only one cavity mode, we have
\begin{equation}
H=\omega b^\dagger b +\sum_{k=1}^N  \omega_k b_k^\dagger b_k + \sum_{k=1}^N C_k (b_k^\dagger+b_k) (b^\dagger+b).
\end{equation}
For the numerical implementation, we rewrite the last equation as  
\begin{eqnarray}
H&=&\frac{p^2}{2M} +\frac{1}{2}M\omega^2 x^2
- \sum_{k=1}^N \tilde{C}_k
x_k x\\
&+&\sum_{k=1}^N \bigg( \frac{p_k^2}{2m_k} +\frac{1}{2}m_k\omega_k^2 x_k^2\bigg) -\sum^N_{k=1} \frac{\hbar \omega_k}{2} - \frac{\hbar \omega}{2}
\end{eqnarray}
with $\tilde{C}_k=C_k \big[\frac{4 m m_k \omega \omega_k}{\hbar^2}\big]^{\frac{1}{2}}$, 
$J(\omega)=\frac{\pi}{2}\sum_{k=1}^N \frac{\tilde{C}_k^2 \delta(\omega-\omega_k)}{m_k \omega_k}$.
To choose the set $\{\tilde{C}_k\}$, we consider that  integrating $J(\omega)$ from 0 to a very large frequency $\Omega$ gives 
$\int_0^\Omega J(\omega)d\omega =\sum_{j=1}^N \frac{\tilde{C}^2_k}{m_k\omega_k}$. Thus, by discretising 
the integral via Riemann sums (the discretisation frequency step is $\Delta$), we get the approximation
\begin{equation}
\sum_{j=1}^N \frac{\tilde{C}^2_k}{m_k\omega_k}= \int_0^\Omega J(\omega)d\omega \approx \sum_{j=1}^N J(\omega_k)\Delta 
\end{equation}
and thus $\frac{\tilde{C}^2_k}{m_k\omega_k}\approx J(\omega_k)\Delta$. In turn, this amounts to say that \cite{Caldeira, Venkataraman}
\begin{equation}
\frac{C^2_k} {m_k\omega_k} \frac{2m_k \omega_k}{\hbar} \frac{2m \omega}{\hbar} \approx J(\omega_k)\Delta \Rightarrow C_k \approx \sqrt{J(\omega_k)}.
\end{equation}
The equation that to implement numerically is 
\begin{equation}
H=\omega b^\dagger b - \sum_{k=1}^N \tilde{C}_k
x_k (b^\dagger+b) 
+\sum_{k=1}^N \bigg( \frac{p_k^2}{2m_k} +\frac{1}{2}m_k\omega_k^2 x_k^2\bigg),
\end{equation}
and thus we must implement $\tilde{C}_k \approx \sqrt{J(\omega_k)}\sqrt{\omega_k}$.

To perform the actual dynamics, we use the quantum-classical (Ehrenfest's) approximation, where the boson system is quantum
but the bath becomes classical. The equations of motion then are:
\begin{eqnarray}
i\frac{d |\psi_{bos}(t)\rangle}{dt}&=&\tilde{H}(\{x_k(t)\}) |\psi_{bos}(t)\rangle\\
m_k \ddot{x}_k(t)&=&-m_k\omega_k^2x_k(t)+C_k(t)\langle b^\dagger +b \rangle_t, \label{Eq_x}\\
\dot{x_k}&=&p_k/m_k \label{Eq_p}\
\end{eqnarray}
where
\begin{eqnarray}
|\psi_{bos}(t)\rangle&=&\sum_{l=0}^M \beta_l(t) ~\frac{(b^\dagger)^l}{\sqrt{l!}} |vac\rangle 
=\sum_{l=0}^M \beta_l(t) ~ |l\rangle \label{bosons}
\\
\tilde{H}(\{x_k(t)\}) &=&\omega b^\dagger b - (b^\dagger+b)\sum_{l=1}^N C_kx_k(t)
\end{eqnarray}
The bosonic Schr\"odinger equation is solved as usual while for the bath fields
we use the coordinate Verlet algorithm. The shape chosen is
$J(\omega_k)\approx A \omega_k^a $, and $C_k\approx \sqrt{k^a}$.\newline

\subsection{Exponential and classical dissipation}\label{APPE}
\begin{figure}
	\begin{center}
	\includegraphics[width=0.6\columnwidth]{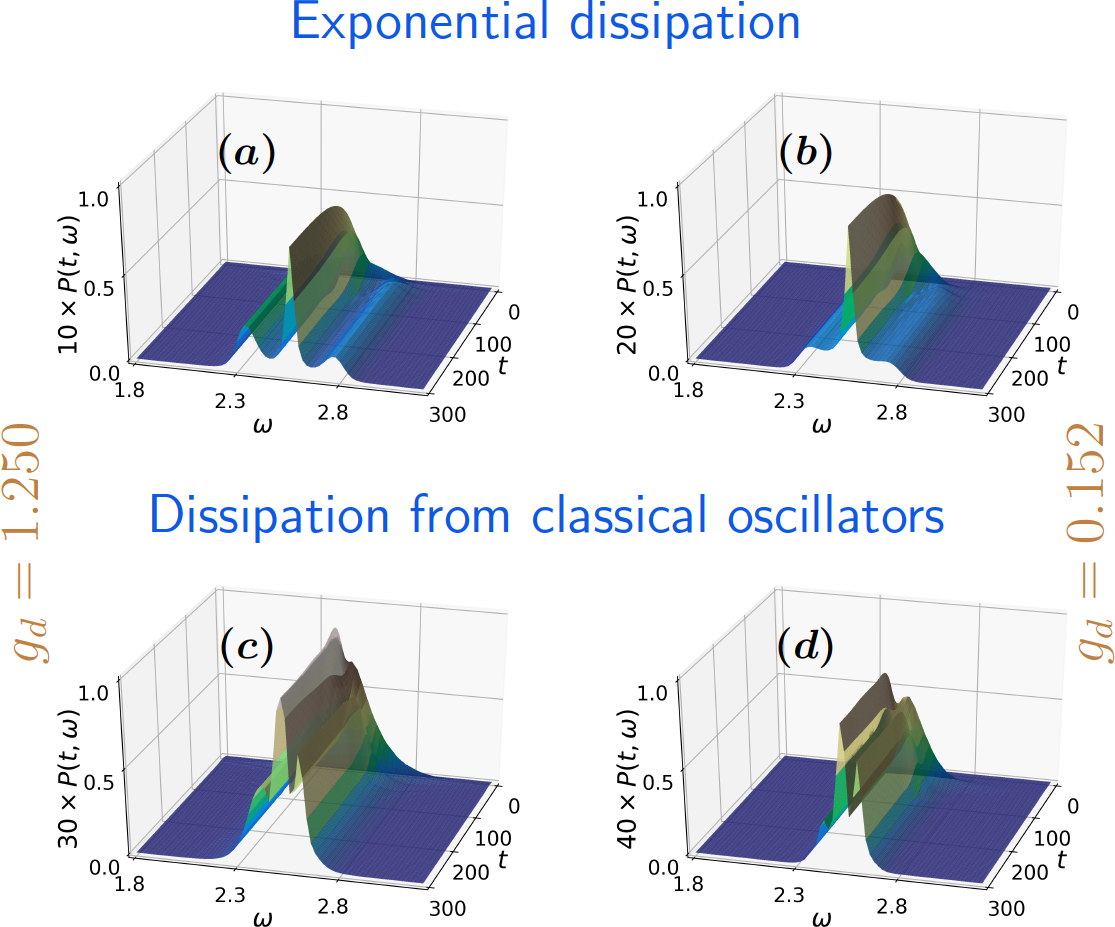}
	\end{center}
	\caption{The calculations corresponds to infinite nuclear mass. The other parameters are $g_{c}=0.03$, $U=1.0$, $g_{f}=0.01$, resonance frequency= $\omega_0=2.56$, and the initial state is $|\Psi_0 ''\rangle$. 	
	The driving is on until $\left< b^\dagger b\right>\approx9$,  $\bm{(a)}$ fast driving with $g_d=1.250$, $t_s=\frac{4\pi}{\omega_0}$ and $\Gamma=0.02$ $\bm{(b)}$ 
	Slow driving with $g_d=0.152$, $t_1=\frac{6\pi}{\omega_0}$, $t_2=\frac{41\pi}{\omega_0}$ and $\Gamma=0.02$ $\bm{(c)}$ fast driving with the classical dissipation (i.e. $\Gamma=0.00$), with $C_k=A(\Delta k)^a$. There are $N_B=$1000 classical oscillators, $A=0.01$, $a=0.8$ and $\Delta k=0.01$. $\bm{(d)}$ Slow driving with the classical dissipation. } 
	\label{figSM2}
\end{figure}

The resonance calculations for the non-rigid molecule in Fig. 4 were performed with the driving field,
which pumps photons in the cavity. The driving was considered 
for dissipation from both exponential and classical-oscillator baths. The results in Fig \ref{figSM2} show the effect of the driving field strength for different form of dissipation. In Fig. \ref{figSM2}a we can observe a Mollow-like spectrum as in the coherent photon case, whereas in Fig. \ref{figSM2}b, the sidebands of the Mollow-like spectrum are less intense.

When dissipation is included via a classical oscillator bath, for both the strength considered, 
the spectrum is not Mollow-like any more, as shown in Fig. \ref{figSM2}c and \ref{figSM2}d.
Since the classical oscillator bath describes the effect of cavity leakage, where the emitted photon disappears faster
from the cavity, the intensity of the corresponding spectrum is less in comparison with the exponential dissipation.        

\bibliographystyle{andp2012}
\providecommand{\WileyBibTextsc}{}
\let\textsc\WileyBibTextsc
\providecommand{\othercit}{}
\providecommand{\jr}[1]{#1}
\providecommand{\etal}{~et~al.}

\end{appendix}


\begin{thebibliography}{99}
\bibitem{semiclassical} This is of no consequence for a rigid molecule, but can have a role in general. We are
currently developing a semiclassical description of the interaction between cavity modes
and nuclear charge and include its effect on the motion of the nuclei.
\bibitem{otherpresc} A different prescription could be to consider, irrespective of the value of $U$,
an incident frequency in resonance with the one particle levels i.e. $\omega_0= 2|V_{eff}|$. Within the perspective adopted here,
this would simply amount to have an off-resonant incident field, with detuning  $\pm |2V_{eff}-\Omega_R|$.
\bibitem{window} We take $\hat{V}_{\rm ext}(t) = g_d (b^\dagger + b) [f(t) \sin \omega_0 t]$, 
with a) $f(t)=\theta(t_s-t )$ a step envelope vanishing after time $t_s$ or b) $f(t)$ a smoothened rectangular pulse. The rectangular pulse $f(t)$ acts approximately between $t_1$ and $t_2$, with envelope
$f(t)=[1-\mathcal{F}_1(t)]\mathcal{F}_2(t)$, where $\mathcal{F}_i(t)=[\exp((t-t_i)/\tau_i)+1]^{-1}$. In all
calculations, $\tau_1=\tau_2=2.0$ whilst the values of $t_1,t_2$ are case specific, and
reported in the figure captions.
\bibitem{footnote} Even with $\beta=3$, the size
of the incident photon subspace $N_i$ must be much larger (explicitly, $N_i=60$) to have good numerical convergence.
\end{thebibliography}

\nolinenumbers

\end{document}